# Topological interfacial states in ferroelectric domain walls of two-dimensional bismuth

Wei Luo[1]†, Yang Zhong[2]†, Hongyu Yu[2], Muting Xie[2], Yingwei Chen[2], Hongjun Xiang[2,3,4]* and Laurent Bellaiche[1]

[1]Physics Department and Institute for Nanoscience and Engineering, University of Arkansas, Fayetteville, Arkansas 72701, USA
[2]Key Laboratory of Computational Physical Sciences (Ministry of Education), State Key Laboratory of Surface Physics, and Department of Physics, Fudan University, Shanghai 200433, China
[3]Collaborative Innovation Center of Advanced Microstructures, Nanjing 210093, China
[4]Shanghai Qi Zhi Institute, Shanghai 200030, China

Corresponding author: hxiang@fudan.edu.cn
†These authors contributed equally to this work.

**Abstract:**

Using machine learning methods, we explore different types of domain walls in the recently unveiled single-element ferroelectric, the bismuth monolayer [Nature 617, 67 (2023)]. Remarkably, our investigation reveals that the charged domain wall configuration exhibits lower energy compared to the uncharged domain wall structure. We also demonstrate that the experimentally discovered tail-to-tail domain wall maintains topological interfacial states caused by the change in the $Z_2$ number between ferroelectric and paraelectric states. Interestingly, due to the intrinsic built-in electric fields in asymmetry DW configurations, we find that the energy of topological interfacial states splits, resulting in an accidental band crossing at the Fermi level. Our study suggests that domain walls in two-dimensional bismuth hold potential as a promising platform for the development of ferroelectric domain wall devices.

**Introduction**

In ferroelectrics (FEs), domain walls (DWs), where the polarization changes direction, play a pivotal role in shaping the overall properties of ferroelectrics. These DWs exhibit unique characteristics distinct from bulk FEs and have been observed to showcase numerous remarkable phenomena [1-7]. One of the most attractive features is their enhanced conductivity [8-12]. Through non-volatile control of DW configurations using an electric field, it becomes possible to manipulate the conductivities of DWs. As a result, DWs can be utilized in the design of various devices, including DW logic, DW transistors, DW diodes, and DW memory[13-22].

Moreover, topological surface states (TSSs), which exist at the surface (or edge for a two-dimensional (2D) system) of topological insulators (TIs), have garnered significant interest in recent years [23-30]. TSSs act as a hallmark at the boundary between trivial insulators ($Z_2=0$) and TIs ($Z_2=1$), theoretically manifesting at DWs. This prompts a natural question: do TSSs exist in FE DWs? If affirmative, it could signify a breakthrough, as TSSs would (i) remain robust irrespective of DW shape distortions, and so forth; (ii) be localized and rapidly decay away from the DW. Additionally, (iii) their relevant electrons would retain high mobilities owing to their Dirac-type band dispersions. To realize such a scenario, a prerequisite is a topological phase transition (note that we refer here to band topology in reciprocal space instead of domain topology in real space) [31-34]) should occur in the polarization switching path. Consequently, two scenarios may unfold. In the first case, the FE phase demonstrates TI behavior, while the paraelectric (PE) phase behaves as a trivial insulator. However, to date, despite predictions of FE TIs [3,35-41], no experimental evidence has been found to support their existence. In the second case, the FE phase behaves as a trivial insulator, while the PE phase exhibits TI behavior. To the best of our knowledge, the recently discovered 2D bismuth (Bi)[42-45] is the candidate to satisfy this latter condition. Hence, investigations of TSSs near DWs in 2D Bi can not only resolve a fundamental physical problem but may also pave the way for high-performance ferroelectric DW devices.

In this study, we utilize ML techniques to explore different DW configurations and electronic properties of 2D Bi. Initially, stable DW configurations are determined using a MLIP. Subsequently, we compute the band structures of different DW configurations employing the ML Hamiltonian approach [46]. Our findings unveil that a charged DW, designated as 1-$xy$ and displaying metallic properties, possesses lower energy compared to an uncharged DW, denoted as 1-$yy$ and characterized as an insulator. Interestingly, we also demonstrate that the experimentally discovered tail-to-tail DW maintains topological interfacial states (TISs), which serve as promising candidates for FE nanoelectronics.

**Results**

**Topological properties of FE and PE for 2D Bi**

The crystal structures of the FE and PE phases of 2D Bi are depicted in Figs. 1a and 1b, with point groups $C_{2v}$ and $D_{2h}$ (space groups: $Pmn2_1$ and $Pmna$), respectively. In the FE state, the polarization is along the -$x$ direction, with a magnitude of approximately $0.4\times10^{-10}$ C/m. The lattice parameters are 4.84 Å (along $x$) and 4.58 Å (along $y$). Band structures for FE and PE states, with and without spin-orbit coupling (SOC) effects, are presented in Fig. S1 of the Supplementary Materials (SM) (Details of calculations are available in the methods section of the SM). Based on these band structures, we speculate that the FE structure is a trivial insulator, while the PE phase is a TI due to band inversion at the $\Gamma$ point (see Figs. S1c and S1d of SM). Notably, for the PE band structure without SOC, we observe four Weyl points (WPs) at general k points in the Brillouin zone (BZ), differing slightly from previous results for 2D Bi [43] (Ref. 37 suggests that WPs locate at high symmetry lines.) This difference originates from the fact that they used the experimental lattice constant, while we use the lattice constant of the FE structure, which is optimized by density functional theory (DFT). This discrepancy indicates that the WPs are accidentally degenerate, a phenomenon highly related to lattice constants or distortions. These four WPs found here are related to each other by time reversal ($T$) and mirror symmetry ($M_y$). Note that the WPs here are protected by space-time symmetry [47-49]. Specifically, the PE structure satisfies $C_{2z}T=UK$, which is an antiunitary operator, where $U$ and $K$ are a unitary matrix and complex conjugation, respectively. By choosing an appropriate basis, we can set $U=1$ since $(C_{2z}T)^2$ is maintained for both spinless and spinful cases[48]. Thus, we have $C_{2z}T=K$. The band crossing Hamiltonian can be expressed as $H(\mathbf{k}) = F_x(\mathbf{k},\delta)\sigma_x + F_y(\mathbf{k},\delta)\sigma_y + F_z(\mathbf{k},\delta)\sigma_z$, where $\delta$ represents an external parameter that can be tuned by lattice constant, pressure, etc. $\sigma_x$, $\sigma_y$ and $\sigma_z$ are the Pauli matrices. The $C_{2z}T=K$ equality necessitates $F_y(\mathbf{k},\delta) = 0$. Therefore, we have $H(\mathbf{k}) = F_x(\mathbf{k},\delta)\sigma_x + F_z(\mathbf{k},\delta)\sigma_z$. Moreover, the band crossing condition requires $F_x = F_z = 0$. In this case, we have three independent variables ($k_x, k_y, \delta$) and two equations. The WPs are thus stable under $C_{2z}T$. With SOC, all WPs open, leading to the non-trivial topological insulator state. To validate our hypothesis, we directly calculated the $Z_2$ in the FE and PE states using the Wilson loop method[50], We found $Z_2=0$ for the FE structure and $Z_2=1$ for the PE phase, as indicated in Figs. S2a and S2b in the SM. Furthermore, we calculated the edge states for the FE and PE structures, and the results are presented in Figs. 1c and 1d. It is evident that the FE structure lacks non-trivial edge states, with the Fermi level (black dashed line) intersecting the edge states an even number of times (specifically, four times in this case). Conversely, for the PE phase, the Fermi level intersects the edge states an odd number of times (three times in this instance), revealing its non-trivial bulk band topology. Consequently, we have demonstrated the existence of a topological phase transition between FE and PE states, substantiated by their maintenance of different topological $Z_2$ numbers. Assuming a DW separates the FE and PE regions, TISs must exist near this DW.

**Stable DW configurations**

Firstly, we verify the reliability of the MLIP. Detailed information is available in part I of the SM. To assess the convergence for DW size, we calculate the DW energy as a function of size. The results indicate that 400 atoms in a supercell are sufficient to describe DW structures. Further details are provided in part II of the SM. Consequently, in the subsequent analysis, the investigated DW structures consistently include 400 atoms in the supercell. Nine initial DW structures are manually constructed and named as 1-$xx$, 1-$xy$, 1-$xz$, 1-$yx$, 1-$yy$, 1-$yz$, $\sqrt{2}$-$xx$, $\sqrt{2}$-$xy$ and $\sqrt{2}$-$xz$. The polarization directions for these structures can be seen in Fig. S4 of the SM. Let's briefly introduce the naming rule for these initial DW structures. Taking 1-$xy$ as an example, the first number "1" indicates that the initial structure for constructing the DW is the unit cell of the FE structure (with 4 atoms in the unit cell). Assuming the polarization is along the +$x$ direction (as indicated by the black arrow in Fig. 2a1), the first letter "$x$" in 1-$xy$ means that we enlarge the cell along the $x$-direction. Specifically, we enlarge the cell 100 times along the $x$-direction, resulting in 400 atoms in the supercell. The second letter "$y$" in 1-$xy$ means that we rotate the right half part by $\pi$ radians along the $y$-axis. After rotation, the polarization is reversed, i.e., along the $-x$ direction (as indicated by the yellow arrow in Fig. 2a1). For $\sqrt{2}$-$xy$, "$\sqrt{2}$" means that the initial structure for constructing the DW is $\sqrt{2}$ times the unit cell of the FE structure (yielding 8 atoms in the cell, named as the $\sqrt{2}$ unit cell). Note that the $\sqrt{2}$ unit cell maintains the same coordinate system as the FE unit cell. Assuming the polarization of the $\sqrt{2}$ unit cell is along the $x$-$y$ direction (as indicated by the black arrow in Fig. 2a4), and since the number of atoms for the investigated DW structure is fixed at 400, we enlarge the cell 50 times along the $x$-direction and rotate the right half parts by $\pi$ radians along the $y$-axis for $\sqrt{2}$-$xy$. After rotation, the polarization is along -$x$ -$y$ (as indicated by the yellow arrow in Fig. 2a4). Similar procedures and notations are applied to distinguish other DW configurations.

Starting from these nine initial DW structures, we optimize them using MLIP implemented in LAMMPS [51]. After optimization, the DW energies, DW widths, and lattice constants changes for each structure are listed in Table 1. Notably, only $\sqrt{2}$-$xx$ and $\sqrt{2}$-$xy$ undergo changes in their DW configurations. Specifically, both evolve into $\sqrt{2}$-$x(x-y)$ (here, $x-y$ indicates a rotation of the right part by $\pi$ radians along the $x-y$ direction). This explains why they essentially have the same DW energy after optimization. From Table 1, it is apparent that 1-$xy$, 1-$yy$ and $\sqrt{2}$-$x(x-y)$ have lower energies. The atomic structures of these three configurations are illustrated in Figs. 2b1, 2b2, and 2b3, respectively. In these figures, the red bar represents the DW. Specifically, 1-$xy$ corresponds to a 180-degree head-to-head charged DW, with the polarization being perpendicular to the DW. On the other hand, 1-$yy$ represents a 180-degree uncharged DW, with the polarization being parallel to the DW. Finally, $\sqrt{2}$-$x(x-y)$ represents a 180-degree charged DW, with the angle between polarization and DW being about 45 degrees. It is worth noting that the 1-$xy$ configuration has been observed experimentally before[42] (note that the head-to-head and tail-to-tail DWs always coexist due to the periodic boundary condition, see Fig. 2b1). As mentioned earlier, 1-$xy$ is a charged DW, while 1-$yy$ is an uncharged DW. Typically, charged DWs have higher energy compared to uncharged (neutral) DWs [52] due to the electrostatic repulsion energy, charged DWs typically have higher energy than uncharged (neutral) DWs. However, in the case of 2D Bi, we observe that the charged DW 1-$xy$ has a lower energy than the uncharged DW 1-$yy$. This anomaly may be attributed to the weak ferroelectricity of single-element ferroelectrics, resulting in a small electrostatic repulsion energy. Given the small energy difference between 1-$xy$ and 1-$yy$ (3.0 meV/atom), we validate this result using DFT. It's important to note that, in these calculations, both configurations now have 80 atoms in the supercell, as it is challenging to calculate properties for 400 atoms within DFT. Our DFT calculations confirm that 1-$xy$ indeed has a lower energy than 1-$yy$, with a difference of about 0.8 meV per atom. In the subsequent discussion, we primarily focus on the low-energy DW configurations 1-$xy$, 1-$yy$, and $\sqrt{2}$-$x(x-y)$, exploring their electronic properties and potential applications.

**Topologically protected interfacial states of DW structures**

We investigate the electronic properties of these DW structures using our ML Hamiltonian. Initially, we validate the reliability of the ML Hamiltonian, with detailed information available in part III of the SM. Building on previous discussions, we anticipate the presence of TISs between FE and PE boundaries due to the change in the $Z_2$ number. To verify that these interfacial states are indeed TISs, we manually construct a DW structure where the left and right regions correspond to PE and FE states (we call it PE-FE configuration), respectively (refer to Fig. 3a). Each region comprises 30-unit cells, corresponding to 14.5 nm. Subsequently, we computed the band structure of this DW structure using the ML Hamiltonian, with the results illustrated in Fig. 3d. A band crossing point (BCP) emerges at the Fermi level. However, this BCP does not correspond to the real Dirac point associated with the TISs, as it does not align with the time-reversal invariant momentum. The underlying physics behind this BCP is depicted in Fig. 3e. Due to the built-in electric field of ferroelectrics (refer to Fig. 3b), the energies of the TISs at two DWs (highlighted by red rectangles in the middle and right in Fig. 3a) split. The TISs with spin-moment locking in the middle DW (indicated by black bands in Fig. 3e) exhibit higher energy compared to those in the right DW (depicted by blue bands in Fig. 3e). This observation is corroborated by the calculations of partial charge density for the first valence band (VB) and first conduction band (CB), as shown in Fig. 3c and Fig. 3b, respectively. It's noteworthy that we also confirm the second valence band (VB-1) and the second conduction band (CB+1) are primarily contributed by the right and middle domain walls, respectively. It's important to note that although the BCP at the Fermi level is not protected by time reversal symmetry, their crossing is unavoidable since Dirac bands of TIs always connect VB and CB.

We note that experimental evidence [42] has observed the tail-to-tail DW structure (due to intrinsic $p$-type doping), containing a PE region with 15 unit cells (about 52 Å). Here, we manually construct the tail-to-tail DW structure with each region comprising 30-unit cells (refer to Fig. 4a1, we call it FE-PE-FE configuration). The calculated band structures from ML are depicted in Fig. 4a. The TISs are identified by green pentagons in Fig. 4a, confirmed by the charge density plot in Fig. 4a1-4a2. In contrast to the scenario in Fig. 3d, the Dirac bands here do not split due to the symmetry of the left and right interfaces (see black lines in Fig. 4a3). Consequently, the Dirac bands exhibit twofold degeneracy (Note: in Fig. 3d, the Dirac bands are singly degenerate), as depicted in Fig. 4a. Each Dirac band originates from two interfaces (black lines in Fig. 4a3) between FE and PE phases. The lower two bands, marked by green hexagons in Fig. 4a, are trivial and result from the interface between FE and FE, as evident from the charge density in Fig. 4a3-a4.

Now, let's examine what occurs in the scenarios of tail-to-tail and head-to-head configurations when the width of the domain wall (paraelectric region) is zero. The structure (1-$xy$, i.e. FE-FE configuration), consisting of 400 atoms, is optimized using our MLIP, and its band structures generated by ML are illustrated in Fig. 4b. Near the Fermi level, there are four bands. The CB and CB+1 (VB and VB-1) originate from the tail-to-tail (head-to-head) domain wall, as evident in Fig. 4b1-b2 (Fig. 4b3-b4). These bands are considered trivial since there is no $Z_2$ jumping near the interfaces (due to the absence of a PE region). Instead, they arise solely from regular charged DWs. The band structures for other DW structures can be viewed in Fig. S7 and Fig. S8 of the SM.

**Conclusions**

In summary, we employed a ML approach based on DFT data to systematically explore stable DW configurations and compute electronic band structures for DW structures of 2D ferroelectric Bi. To the best of our knowledge, this is the first instance where the ML Hamiltonian has been utilized to investigate TISs for such a large system. Our investigation reveals that the charged DW exhibits lower energy compared to the uncharged DW. Additionally, the ML Hamiltonian-predicted band structures confirm the existence of TISs at the interface between FE and PE phases. Furthermore, we directly observe the splitting of TSSs in asymmetric ferroelectric DWs due to built-in electric fields. We anticipate that these discoveries will offer significant advantages for applications in nanoelectronics relying on non-trivial low-dimensional ferroelectrics.

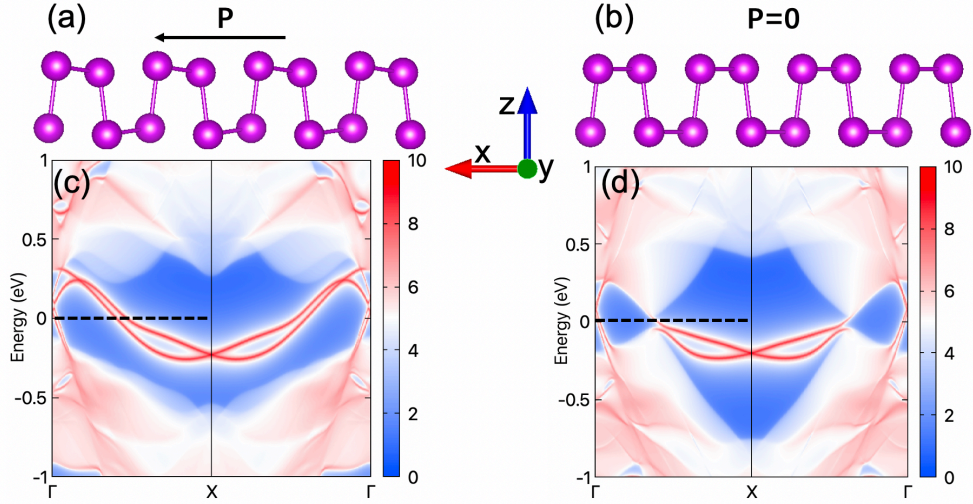

Fig. 1 Side view of FE (a) and PE (b) structures. The black arrow represents the direction of polarization. Edge states for FE (c) and PE (d) states. The black dashed line indicates the Fermi level. The red (blue) color represents higher (lower) local density of states (LDOS). The surface states can be clearly seen as red lines.

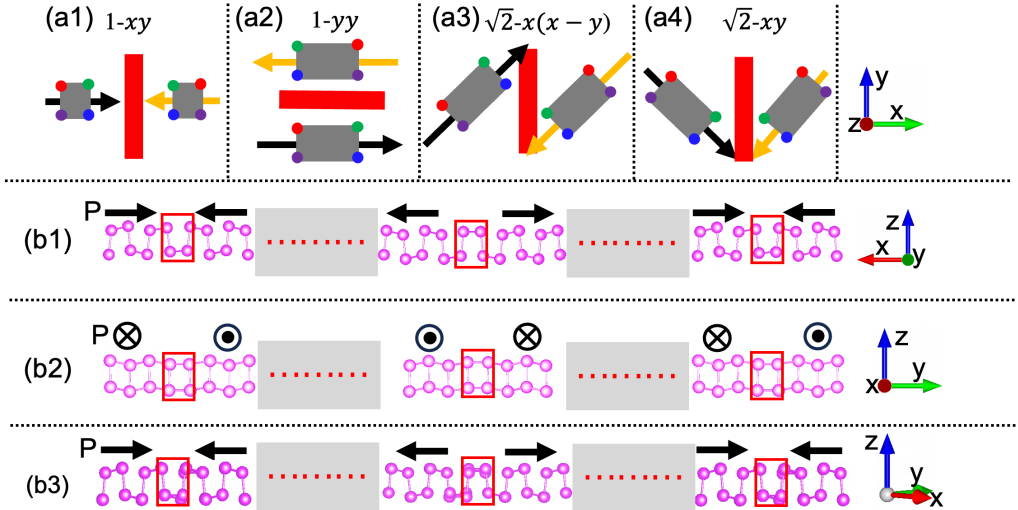

Fig. 2 Schematic diagrams describing the DW structures: 1-$xy$ (a1), 1-$yy$ (a2), $\sqrt{2}$-$x(x-y)$ (a3), and $\sqrt{2}$-$xy$ (a4). The red bar represents the DW, and black and yellow arrows show polarization directions. Bi atoms are represented by red, blue, green, and purple dots. Atomic structures corresponding to 1-$xy$ (b1), 1-$yy$ (b2), and $\sqrt{2}$-$x(x-$

*y*) (b3) are shown, with the red square indicating the DWs. Note that a4 represents the initial configuration, which will change to a3 after structure optimization.

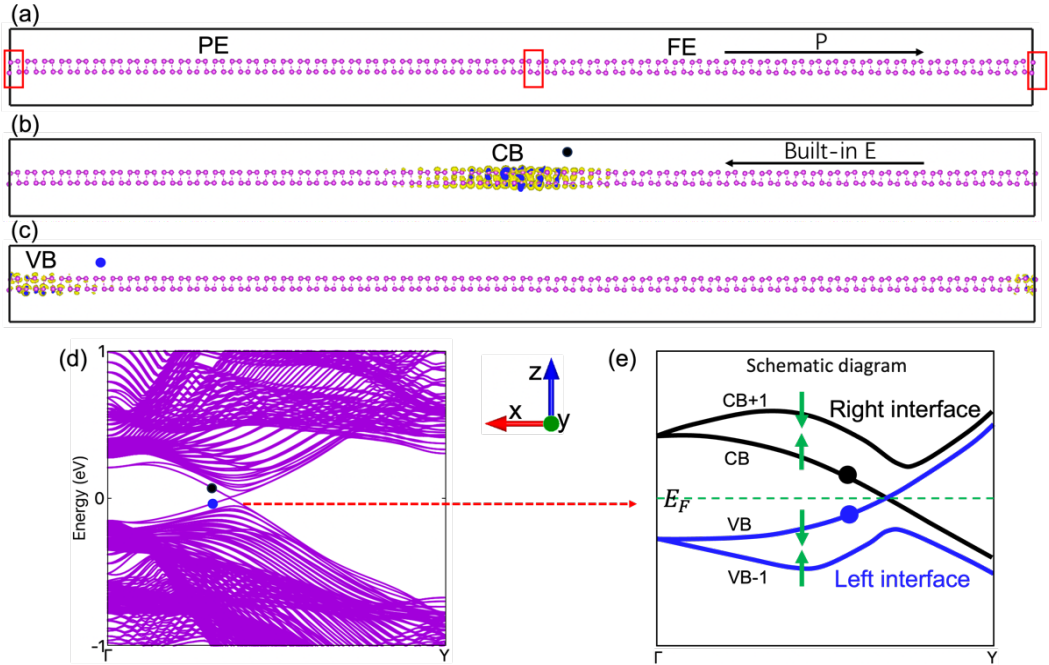

Fig. 3 (a) Crystal structure of the PE-FE configuration. The red squares depict the domain DWs. Charge density for the CB (b) and VB (c) near the BCP. (d) Band structures for the PE-FE configuration. (e) The schematic diagram for the topological interfacial bands. Green arrows indicate spin directions.

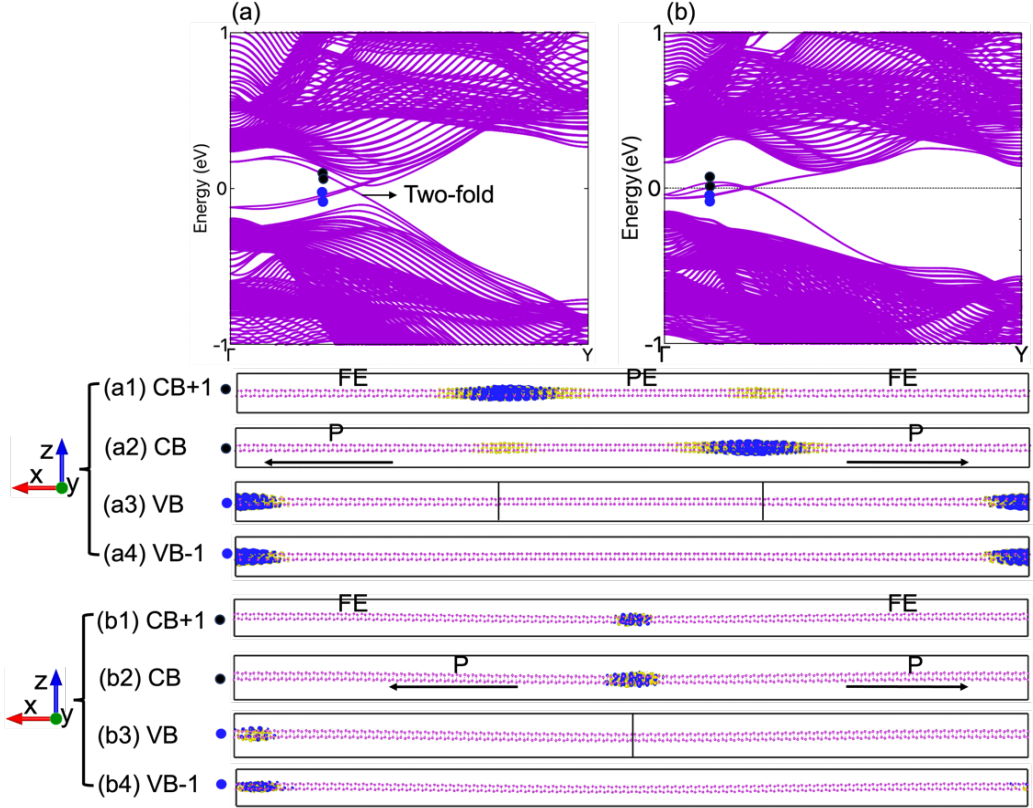

Fig. 4 Band structures generated by ML for the FE-PE-FE (a) and FE-FE (b) configurations. The charge density distribution of the CB (CB+1) and VB (VB-1) for FE-PE-FE (a1-a4) and FE-FE (b1-b4) obtained from ML calculations.

Table 1: DW energies, DW widths, and lattice constants changes for the nine considered DW structures after structural optimization. The red text indicates that $\sqrt{2}\text{-}xx$ and $\sqrt{2}\text{-}xy$ evolve into $\sqrt{2}\text{-}x(x-y)$ after structural optimization. Note that $1\text{-}xy$, $1\text{-}yy$, and $\sqrt{2}\text{-}x(x-y)$ are low-energy configurations discussed in Fig. 2, while the other six configurations are high-energy configurations. All nine of these considered DW structures can be seen in Fig. S4 of the SM

| Configuration | $1\text{-}xx$ | $1\text{-}xy$ | $1\text{-}xz$ | $1\text{-}yx$ | $1\text{-}yy$ | $1\text{-}yz$ | $\sqrt{2}\text{-}xx$ [$\sqrt{2}\text{-}x(x-y)$] | $\sqrt{2}\text{-}xy$ [$\sqrt{2}\text{-}x(x-y)$] | $\sqrt{2}\text{-}xz$ |
|---|---|---|---|---|---|---|---|---|---|
| Formation energy(meV) | 515.8 | 217.8 | 484.8 | 670.8 | 220.8 | 581.8 | 266.8 | 266.7 | 540.8 |
| Domain wall width | 1 UC | 1 UC | 1 UC | 1 UC | 1 UC | 1 UC | 4 UC | 4 UC | 12 UC |
| Lattice constant change ($x$) | -2.9% | -2.6% | -3.0% | -3.1% | -2.6% | -3.0% | -5.2% | -5.2% | -5.6% |
| Lattice constant change ($y$) | -2.7% | -2.9% | -2.7% | -2.5% | -2.8% | -2.7% | -0.1% | -0.1% | 0.1% |

**Acknowledgements**

The work at Fudan is supported by the National Key R&D Program of China (No. 2022YFA1402901), NSFC (grants No. 11825403, 11991061, 12188101), and the Guangdong Major Project of the Basic and Applied Basic Research (Future functional materials under extreme conditions--2021B0301030005). W. L. and B. L. thank the support of Vannevar Bush Faculty Fellowship (VBFF) grant no. N00014-20-1-2834 from the Department of Defense. Y.Z., H.Y., M.X., Y.C. and H.X. were supported by NSFC (Grants No. 11825403, 11991061 and No. 12188101). The Arkansas High Performance Computing Center (AHPCC) is also acknowledged.

**References**

[1] W. Jiang, M. Noman, Y. Lu, J. Bain, P. Salvador, and M. Skowronski, Journal of Applied Physics **110** (2011).
[2] S. Liu, F. Zheng, N. Z. Koocher, H. Takenaka, F. Wang, and A. M. Rappe, The journal of physical chemistry letters **6**, 693 (2015).
[3] S. Liu, Y. Kim, L. Z. Tan, and A. M. Rappe, Nano Letters **16**, 1663 (2016).
[4] M. Calleja, M. T. Dove, and E. K. Salje, Journal of Physics: Condensed Matter **15**, 2301 (2003).
[5] Y. Geng, H. Das, A. L. Wysocki, X. Wang, S. Cheong, M. Mostovoy, C. J. Fennie, and W. Wu, Nature materials **13**, 163 (2014).
[6] M. Matsubara *et al.*, Science **348**, 1112 (2015).
[7] M. Giraldo, Q. N. Meier, A. Bortis, D. Nowak, N. A. Spaldin, M. Fiebig, M. C. Weber, and T. Lottermoser, Nature communications **12**, 3093 (2021).
[8] R. Vasudevan *et al.*, Nano letters **12**, 5524 (2012).
[9] R. K. Vasudevan *et al.*, Advanced Functional Materials **23**, 2592 (2013).
[10] C. Godau, T. Kämpfe, A. Thiessen, L. M. Eng, and A. Haußmann, ACS nano **11**, 4816 (2017).
[11] J. Seidel *et al.*, Nature materials **8**, 229 (2009).
[12] G. Catalan, J. Seidel, R. Ramesh, and J. F. Scott, Reviews of Modern Physics **84**, 119 (2012).
[13] J. Jiang *et al.*, Nature materials **17**, 49 (2018).
[14] M. Campbell, J. McConville, R. McQuaid, D. Prabhakaran, A. Kumar, and J. Gregg, Nature Communications **7**, 13764 (2016).
[15] P. W. Turner, J. P. McConville, S. J. McCartan, M. H. Campbell, J. Schaab, R. G. McQuaid, A. Kumar, and J. M. Gregg, Nano letters **18**, 6381 (2018).
[16] P. Sharma, Q. Zhang, D. Sando, C. H. Lei, Y. Liu, J. Li, V. Nagarajan, and J. Seidel, Science advances **3**, e1700512 (2017).
[17] J. Whyte and J. Gregg, Nature Communications **6**, 7361 (2015).
[18] G. Nataf, M. Guennou, J. Gregg, D. Meier, J. Hlinka, E. Salje, and J. Kreisel,

Nature Reviews Physics **2**, 634 (2020).
[19] D. Meier and S. M. Selbach, Nature Reviews Materials **7**, 157 (2022).
[20] H. Sun *et al.*, Nature communications **13**, 4332 (2022).
[21] B. Vul, G. Guro, and I. Ivanchik, Ferroelectrics **6**, 29 (1973).
[22] X. Chai, J. Jiang, Q. Zhang, X. Hou, F. Meng, J. Wang, L. Gu, D. W. Zhang, and A. Q. Jiang, Nature communications **11**, 2811 (2020).
[23] M. Z. Hasan and C. L. Kane, Reviews of modern physics **82**, 3045 (2010).
[24] F. Zhang, C. L. Kane, and E. J. Mele, Physical Review B **86**, 081303 (2012).
[25] J. E. Moore, Nature **464**, 194 (2010).
[26] X.-L. Qi and S.-C. Zhang, Reviews of Modern Physics **83**, 1057 (2011).
[27] L. Fu, C. L. Kane, and E. J. Mele, Physical review letters **98**, 106803 (2007).
[28] W. Luo and H. Xiang, Nano letters **15**, 3230 (2015).
[29] C.-C. Liu, W. Feng, and Y. Yao, Physical review letters **107**, 076802 (2011).
[30] I. Knez, R.-R. Du, and G. Sullivan, Physical review letters **107**, 136603 (2011).
[31] S. Chae, Y. Horibe, D. Jeong, N. Lee, K. Iida, M. Tanimura, and S.-W. Cheong, Physical review letters **110**, 167601 (2013).
[32] F.-T. Huang *et al.*, Nature communications **7**, 11602 (2016).
[33] J. Junquera *et al.*, Reviews of Modern Physics **95**, 025001 (2023).
[34] W. Luo, A. Akbarzadeh, Y. Nahas, S. Prokhorenko, and L. Bellaiche, Nature Communications **14**, 7874 (2023).
[35] L. Kou, H. Fu, Y. Ma, B. Yan, T. Liao, A. Du, and C. Chen, Physical Review B **97**, 075429 (2018).
[36] B. Monserrat, J. W. Bennett, K. M. Rabe, and D. Vanderbilt, Physical review letters **119**, 036802 (2017).
[37] D. Di Sante, P. Barone, A. Stroppa, K. F. Garrity, D. Vanderbilt, and S. Picozzi, Physical review letters **117**, 076401 (2016).
[38] J. He, D. Di Sante, R. Li, X.-Q. Chen, J. M. Rondinelli, and C. Franchini, Nature communications **9**, 1 (2018).
[39] M. U. Rehman, C. Hua, and Y. Lu, Chinese Physics B **29**, 057304 (2020).
[40] J. Liu, T. H. Hsieh, P. Wei, W. Duan, J. Moodera, and L. Fu, Nature materials **13**, 178 (2014).
[41] T. H. Hsieh, H. Lin, J. Liu, W. Duan, A. Bansil, and L. Fu, Nature communications **3**, 982 (2012).
[42] J. Gou *et al.*, Nature **617**, 67 (2023).
[43] Y. Lu *et al.*, Nano letters **15**, 80 (2015).
[44] C. Xiao, F. Wang, S. A. Yang, Y. Lu, Y. Feng, and S. Zhang, Advanced Functional Materials **28**, 1707383 (2018).
[45] S. Zhong, X. Zhang, S. Liu, S. A. Yang, and Y. Lu, Physical Review Letters **131**, 236801 (2023).
[46] Y. Zhong, H. Yu, M. Su, X. Gong, and H. Xiang, arXiv preprint arXiv:2210.16190 (2022).
[47] C. Fang and L. Fu, Physical Review B **91**, 161105 (2015).
[48] J. Ahn and B.-J. Yang, Physical Review Letters **118**, 156401 (2017).
[49] W. Luo, J. Ji, J. Lu, X. Zhang, and H. Xiang, Physical Review B **101**, 195111

(2020).

[50] R. Yu, X. L. Qi, A. Bernevig, Z. Fang, and X. Dai, Physical Review B **84**, 075119 (2011).

[51] A. P. Thompson *et al.*, Computer Physics Communications **271**, 108171 (2022).

[52] M. Y. Gureev, A. K. Tagantsev, and N. Setter, Physical Review B **83**, 184104 (2011).